\documentclass[12pt,a4paper,twoside]{article}

\usepackage{authblk}

\usepackage[utf8]{inputenc}
\usepackage[T1]{fontenc}
\usepackage{libertine}
\usepackage{microtype} 

\usepackage{mathrsfs}
\usepackage[english]{babel} 
\usepackage{bbm}
\usepackage{amsmath,amsfonts,amscd,amssymb,latexsym}

\usepackage{graphics}
\graphicspath{{/}}
\DeclareGraphicsExtensions{.pdf,.png,.jpg}

\usepackage{rotating}
\usepackage{longtable}
\usepackage{multirow}
\usepackage{cite}
\usepackage{geometry} 
\usepackage[hang, small,labelfont=bf,up,textfont=up]{caption} 
\usepackage{booktabs} 
\usepackage{float} 
\usepackage{hyperref} 

\usepackage{lettrine} 
\usepackage{paralist} 

\usepackage{abstract} 

\usepackage{titlesec} 
\titleformat{\section}[block]{\large\scshape}{\thesection.}{1em}{} 
\titleformat{\subsection}[block]{\large}{\thesubsection.}{1em}{} 

\usepackage{fancyhdr} 
\newcommand{\z}{&&\hspace*{-1cm}}

\newcommand{\bea}{\begin{eqnarray}}
\newcommand{\eea}{\end{eqnarray}}
\newcommand{\be}{\begin{equation}}
\newcommand{\ee}{\end{equation}}

\newcommand{\ar}{a_s}

\title{\fontsize{16pt}{12pt}\selectfont\textbf{About Fractional Analytic QCD  beyond Leading Order}} 

\author[1]{A.V.~Kotikov\thanks{Corresponding Author: kotikov@theor.jinr.ru}}
\author[1,2]{I.A.~Zemlyakov}
\affil[1]{\small{Bogoliubov Laboratory of Theoretical Physics, Joint Institute for Nuclear Research,
    141980 Dubna, Russia}}
\affil[2]{Dubna State University,
Dubna, Moscow Region, Russia}

\date{\today}
\begin{document}
\maketitle 
\thispagestyle{fancy} 


\begin{abstract}
  We present an overview of fractional analytic QCD beyond leading order, following the results recently obtained in Ref. \cite{Kotikov:2022sos}.
  We demonstrate four different representations, the details of their derivation, and show the applicability of analytic
  QCD to the analysis of the Bjorken sum rule.\\
\end{abstract}
\hrule


\section{ Introduction }

According to the general principles of (local) quantum field theory (QFT) \cite{Bogolyubov:1959bfo,Oehme:1994pv},
observables in the spacelike region can have singularities only for negative values of their argument $Q^2$.
For large values of $Q^2$, however, these observables are usually represented as power expansions in the running coupling constant
$\alpha_s(Q^2)$,
which has a ghostly singularity, the so-called Landau pole, at $Q^2 = \Lambda^2$. Therefore, to restore the analyticity of the expansions under consideration,
this pole in the strong coupling constant should be removed.

Indeed, the strong coupling constant $\alpha_s(Q^2)$ obeys  the renormalization group equation
\be
L\equiv \ln\frac{Q^2}{\Lambda^2} = \int^{\overline{a}_s(Q^2)} \, \frac{da}{\beta(a)},~~ \overline{a}_s(Q^2)=\frac{\alpha_s(Q^2)}{4\pi}\,
\label{RenGro}
\ee
with some boundary condition and the QCD $\beta$-function:
\be
\beta(\ar) ~=~ -\sum_{i=0} \beta_i \overline{a}_s^{i+2} 
=-\beta_0  \overline{a}_s^{2} \, \Bigl(1+\sum_{i=1} b_i \ar^i \Bigr),~~ b_i=\frac{\beta_i}{\beta_0^{i+1}}\,, ~~
\ar(Q^2)=
\beta_0\,\overline{a}_s(Q^2)\,, 
\label{beta}
\ee
where
\be
\beta_0=11-\frac{2f}{3},~~\beta_1=102-\frac{38f}{3},~~\beta_2=\frac{2857}{2}-\frac{5033f}{18}+\frac{325f^2}{54},~~
\label{beta_i}
\ee
for $f$ active quark flavors. Really now the first fifth coefficients, i.e. $\beta_i$ with $i\leq 4$, are exactly known \cite{Baikov:2016tgj,Herzog:2017ohr,Luthe:2017ttg}.
In our present consideration we will need only $0 \leq i\leq 2$.

We note that in Eq. (\ref{beta})
we added the first coefficient of the  QCD $\beta$-function to the $\ar$ definition, as is usually the case  in the case of  
of analytic coupling constants (see, e.g., Refs. \cite{ShS}-\cite{Ayala:2018ifo}).

So, already at leading order (LO), when $\ar(Q^2)=\ar^{(1)}(Q^2)$, we have from Eq. (\ref{RenGro})
\be
\ar^{(1)}(Q^2) = \frac{1}{L}\, ,
\label{asLO}
\ee
i.e. $\ar^{(1)}(Q^2)$ does contain a pole at $Q^2=\Lambda^2$.

In a series of papers \cite{ShS,MSS,Sh} by Shirkov and Solovtsov an effective approach was developed to eliminate the Landau singularity without
introducing extraneous infrared (IR) controllers, such as the effective gluon mass
(see, for example, \cite{Parisi:1979se,Cornwall:1981zr,GayDucati:1993fn,Mattingly:1992ud}).

This method is based on the dispersion relation, which connects the new analytic coupling constant $A_{\rm MA}(Q^2)$ with the spectral function $r_{\rm pt}(s)$
obtained in the framework of perturbation theory. In LO this gives
    \be
A^{(1)}_{\rm MA}(Q^2) 
= \frac{1}{\pi} \int_{0}^{+\infty} \, 
\frac{ d s }{(s + t)} \, r^{(1)}_{\rm pt}(s) \, ,
\label{disp_MA_LO}
\ee
where
\be
r^{(1)}_{\rm pt}(s)= {\rm Im} \; a_s^{(1)}(-s - i \epsilon) \,.
\label{SpeFun_LO}
\ee
The \cite{ShS,MSS,Sh} approach follows the corresponding results \cite{Bogolyubov:1959vck} obtained in the framework of Quantum Electrodynamics.

So, we want to repeat once again: the spectral function in the dispersion relation (\ref{disp_MA_LO}) is taken directly from perturbation theory, and the analytical
coupling constant $A_{\rm MA}(Q^2)$ is restored using this dispersion relation. This approach is usually called the {\it Minimal Approach} (MA) (see, for example, \cite{Cvetic:2008bn})
or {\it Analytical Perturbation Theory} (APT) \cite{ShS,MSS,Sh}.
\footnote{An overview of other similar approaches can be found in \cite{Bakulev:2008td}, including approaches \cite{Nesterenko:2003xb} that are close to APT.}

Thus, MA QCD is a very convenient approach that combines the analytical properties of QFT quantities and the results
obtained in the framework of perturbative QCD, leading to the appearance of the MA coupling constant $A_{\rm MA}(Q^2 )$, which is close to the usual strong coupling constant
$a_s(Q^2)$ in the limit of large values of its argument $Q^2$ and completely different from $a_s(Q^2)$ for small values of its argument $Q^2$,
i.e. for $Q^2 \leq \Lambda^2$.

A further development of APT is the so-called fractional APT (FAPT), which extends the construction principles described above to perturbative series
starting from non-integer powers of the coupling constant. Within the QFT framework, such series arise for quantities having non-zero anomalous dimensions
(see the famous papers \cite{BMS1,Bakulev:2006ex,Bakulev:2010gm}, some  privious study \cite{Karanikas:2001cs} and reviews in \cite{Bakulev:2008td}).
Compact expressions for quantities in the FAPT framework were obtained mainly at the LO, however, the FATF approach was also used in higher orders,
mainly by re-expanding the corresponding coupling constants in powers of the LO coupling constant, as well as using some approximations.

In this paper, we give an overview of the main properties of higher-order MA coupling constants in the framework of FAPT, obtained in Ref.  \cite{Kotikov:2022sos}
using the so-called $1/L$-expansion. Note that for an ordinary coupling constant, this expansion is applicable only for large values of its argument $Q^2$,
i.e. for $Q^2>>\Lambda^2$; however, as was shown in  \cite{Kotikov:2022sos}, in the case of an analytic coupling constant, the situation is very different,
and this $1/L$-expansion is applicable for all
values of the argument. This is due to the fact that the non-leading expansion corrections vanish not only at $Q^2 \to \infty$, but also at $Q^2 \to 0$,
\footnote{The absence of high-order corrections for $Q^2 \to 0$ was also discussed in Refs. \cite{ShS,MSS,Sh}.}
which leads only to nonzero (small) corrections in the region $Q^2 \sim \Lambda^2$ (see Section 4 below for a detailed discussion).

Below we consider four different representations for the MA coupling constant and its (fractional) derivatives obtained in  \cite{Kotikov:2022sos} and valid in
principle in any order of perturbation theory. However, in order to avoid cumbersome formulas, but at the same time to show the main features of the approach
obtained in  \cite{Kotikov:2022sos}, we restrict ourselves to considering only the first three orders of perturbation theory.

The paper is organized as follows. In Section 2, we first consider the main properties of the usual strong coupling constant and its $1/L$-decomposition.
Section 3 contains fractional derivatives (that is, $\nu$-derivatives) of the usual strong coupling constant, $1/L$-expansions of which can be represented as
the application of some operators acting on the $\nu$-derivatives of the LO strong coupling constant. This is the key idea of Ref. \cite{Kotikov:2022sos}, which makes
it possible
to construct $1/L$-expansions of the $\nu$-derivatives of the MA-coupling constant at high-order perturbation theory, two different possibilities of which are
presented in Sections 4 and 5. In addition, in Section 6, two integral representations of the $\nu$-derivatives of the MA-coupling constant at high-order perturbation theory,
are presented. One is based on the spectral density obtained in high orders of perturbation theory, and the other is obtained using the above operators.
Section 7 contains an application of this approach to the Bjorken sum rule. Finally, some final discussions are presented. In addition, we have several applications.
Appendix A contains formulas for restoring the $\nu$-derivatives of strong coupling constant at higher orders.
Appendix B presents some alternative results for $\nu$-derivatives of the MA coupling constant that may be useful for some applications.

\section{Strong coupling constant}
\label{strong}

As shown in Introduction, the strong coupling constant $a_s(Q^2)$ obeys the renormalized group equation (\ref{RenGro}).
When $Q^2>>\Lambda^2$, the Eq. (\ref{RenGro})
can be solved by iterations in the form of a $1/L$-expansion
(we give the first three terms of the expansion in accordance with the reasoning in the introduction),
which can be represented in the following compact form
\be
a^{(1)}_{s,0}(Q^2) = \frac{1}{L_0},~~
a^{(i+1)}_{s,i}(Q^2) = 
a^{(1)}_{s,i}(Q^2) + \sum_{m=2}^i \, \delta^{(m)}_{s,i}(Q^2)
\,,~~(i=0,1,2,...)\,,
\label{as}
\ee
where
\be
L_k=\ln t_k,~~t_k=\frac{1}{z_k}=\frac{Q^2}{\Lambda_k^2}\,.
\label{L}
\ee

The corrections $\delta^{(m)}_{s,k}(Q^2)$ are represented as follows
\be
\delta^{(2)}_{s,k}(Q^2) = - \frac{b_1\ln L_k}{L_k^2} ,~~
\delta^{(3)}_{s,k}(Q^2) =  \frac{1}{L_k^3} \, \Bigl[b_1^2(\ln^2 L_k-\ln L_k-1)+b_2\Bigr]\,.
\label{ds}
\ee

It is shown in the equations (\ref{as}) and (\ref{ds}) that in any perturbation order, the coplant $\ar(Q^2)$ contains its own dimensional transmutation parameter
$\Lambda$, which is fixed by comparing with experimental results. This parameter is related to the normalization of $\alpha_s(M_Z^2)$ as
\be
\Lambda_{i}=M_Z \, \exp\left\{-\frac{1}{2} \left[\frac{1}{a_s(M_Z^2)} + b_1\, \ln a_s(M_Z^2) +
\int^{\overline{a}_s(M_Z^2)}_0 \, da \, \left(\frac{1}{\beta(a)}+ \frac{1}{a^2(\beta_0+\beta_1 a)}\right)\right]\right\}\,,
\label{Lambdai}
\ee
where $\alpha_s(M_Z)=0.1176$ in  PDG20 \cite{PDG20}.

\subsection{$f$-dependence of the coupling constant $\ar(Q^2)$ }

The coefficients $\beta_i$ depend on the number $f$ of active quarks
(see Eq. (\ref{beta_i})) that change the coupling constant $\ar(Q^2)$ at $Q^2_f \sim m^2_f$ thresholds, where some extra quark comes into play
at $Q^2 > Q^2_f$. Here $m_f$ is the mass $\overline{MS}$ of the $f$ quark, for example,
$m_b=4.18 +0.003-0.002$ GeV and $m_c=1.27 \pm 0.02$ GeV from PDG20 \cite{PDG20}.
\footnote{Strictly speaking, the quark masses in the $\overline{MS}$ scheme depend on $Q^2$ and $m_f=m_f(Q^2=m_f^2)$. The $Q^2$-dependence is rather slow and
  will not be discussed in this study.}
Thus the coupling constant $a_s$ depends on $f$, and this $f$-dependence can be taken into account in $\Lambda$, i.e. it is $\Lambda^f$ that contributes to the above
Eqs. (\ref{RenGro}) and (\ref{as}).
Moreover, Eq. (\ref{Lambdai}) can indeed be used to determine $\Lambda_{i}^{f=5}$, since five quarks are active at $Q^2=M_Z^2$.

Relationships between $\Lambda_{i}^{f}$ and $\Lambda_{i}^{f-1}$ can be obtained from Eq. (\ref{Lambdai}) with the replacement
$M_Z \to Q_f$ and the so-called matching conditions, i.e. links between $a_s(f,Q_f^2)$ and $a_s(f-1,Q_f^2)$. In the $\overline{MS}$ scheme,
the matching conditions are known up to the four-loop order \cite{Chetyrkin:2005ia} and they are usually used
for $Q_f^2=m_f^2$, where these relations have the most a simple view (see e.g. \cite{FLAG,Enterria} for a recent review).

Here we will not consider the $f$-dependence of $\Lambda_{i}^{f}$ and $a_s(f,M_Z^2)$, since we are mainly considering the region of small $Q^2$. We use the results
for $\Lambda_{i}^{f=3}$, which we need to construct an analytic coupling constant for small values of $Q^2$.

\begin{figure}[!htb]
\centering
\includegraphics[width=0.58\textwidth]{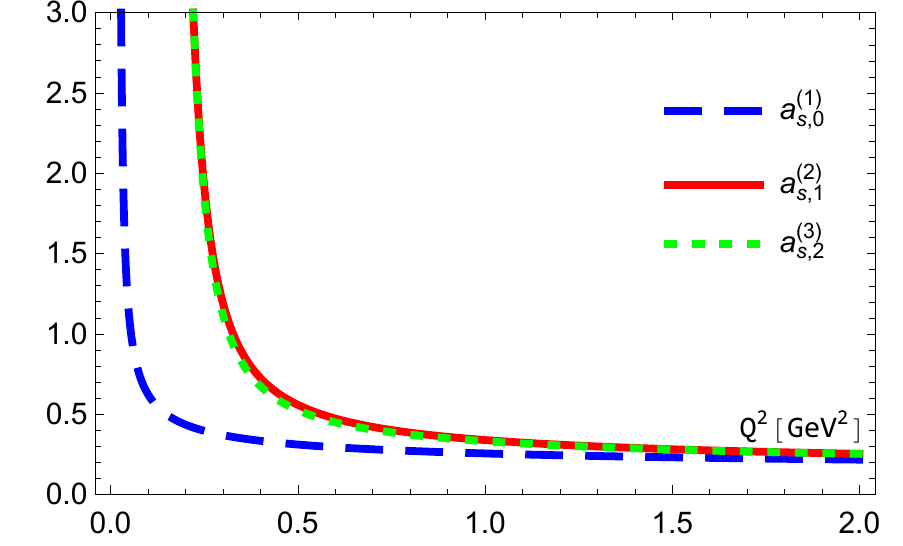}
\caption{\label{fig:as1352}
        The results for $a^{(i+1)}_{s,i}(Q^2)$ with $i=0,1,2$. Here and in the following figures, the $\Lambda_i^{f=3}$ values shown in (\ref{Lambdas}) are used.}
\end{figure}

On Fig. \ref{fig:as1352} one can see that the strong coupling constants $a^{(i+1)}_{s,i}(Q^2)$ become singular at $Q^2=\Lambda_i^2$.
The values of $\Lambda_0$ and $\Lambda_i$ $(i\geq 1)$ are very different.
We use their results taken from a recent Ref. \cite{Chen:2021tjz}, where $\Lambda_i^{f=3}$
$(i=0,1,2)$ were obtained in the following form
\be
\Lambda_0^{f=3}=142~~ \mbox{MeV},~~\Lambda_1^{f=3}=367~~ \mbox{MeV},~~\Lambda_2^{f=3}=324~~ \mbox{MeV}
\,.
\label{Lambdas}
\ee

\section{Fractional derivatives}

Following \cite{Cvetic:2006mk,Cvetic:2006gc},
we introduce the derivatives (in the $(i+1)$-order of perturbation theory)
\be
\tilde{a}^{(i+1)}_{n+1}(Q^2)=\frac{(-1)^n}{n!} \, \frac{d^n a^{(i+1)}_s(Q^2)}{(dL)^n} \, ,
\label{tan+1}
\ee
which are very convenient in the case of the analytical QCD.

The series of derivatives $\tilde{a}_{n}(Q^2)$ can successfully replace the corresponding series of $\ar$-degrees. Indeed, each
the derivative reduces the $\ar$ degree, but is accompanied by an additional $\beta$-function $\sim \ar^2$.
Thus, each application of a derivative produces an additional $\ar$, and thus one can indeed use series of derivatives instead of
series of $\ar$-powers.

In LO, the series of derivatives $\tilde{a}_{n}(Q^2)$ are exactly the same as $\ar^{n}$. Beyond LO, the relationship between $\tilde{a}_{n}(Q^2)$
and $\ar^{n}$ was established in \cite{Cvetic:2006gc,Cvetic:2010di} and extended to the fractional case, where $n \to$ is a non-integer $\nu $, in Ref.
\cite{GCAK}.

Now consider the $1/L$-expansion of $\tilde{a}^{(k)}_{\nu}(Q^2)$. We can raise the $\nu$-power of the results (\ref{as}) and (\ref{ds}) and then
recover $\tilde{a}^{(k)}_{ \nu}(Q^ 2)$ using the relations between $\tilde{a}_{\nu}$ and $\ar^{\nu}$ obtained in \cite{GCAK}.
This operation is carried out in detail in Appendix A. Here we present only the final results, which have the following form
\footnote{The expansion (\ref{tdmp1N}) is very similar to those used in Refs. \cite{BMS1,Bakulev:2006ex} for the expansion of
  ${\bigl({a}^{(i+1)} _{s,i}(Q^2)\bigr)}^ {\nu}$ in terms of powers of $a^{(1)}_{s,i}(Q^2)$.}:
\bea
&&\tilde{a}^{(1)}_{\nu,0}(Q^2)={\bigl(a^{(1)}_{s,0}(Q^2)\bigr)}^{\nu} = \frac{1}{L_0^{\nu}},~~
\tilde{a}^{(i+1)}_{\nu,i}(Q^2)=\tilde{a}^{(1)}_{\nu,i}(Q^2) + \sum_{m=1}^{i}\, C_m^{\nu+m}\, \tilde{\delta}^{(m+1)}_{\nu,i}(Q^2),~~\nonumber\\
&&\tilde{\delta}^{(m+1)}_{\nu,i}(Q^2)=
\hat{R}_m \, \frac{1}{L_i^{\nu+m}},~~C_m^{\nu+m}=\frac{\Gamma(\nu+m)}{m!\Gamma(\nu)}\,,
\label{tdmp1N}
\eea
where
\bea
&&\hat{R}_1=b_1 \Bigl[\hat{Z}_1(\nu)+ \frac{d}{d\nu}\Bigr],~~
\hat{R}_2=b_2 + b_1^2 \Bigl[\frac{d^2}{(d\nu)^2} +2 \hat{Z}_1(\nu+1)\frac{d}{d\nu} + \hat{Z}_2(\nu+1 )\Bigr]
\, .
\label{hR_i}
\eea

The representation (\ref{tdmp1N}) of the $\tilde{\delta}^{(m+1)}_{\nu,i}(Q^2)$ corrections in the form of $\hat{R} _m$-operators is very important and allows us
to present high-order results for the ($1/L$-expansion) of an analytic coupling constant in a similar way (see the next Section).

Note that, using rather complex forms for the coplant powers ${[a_s^{(i+1)}(Q^2)]}^{\nu}$ and for the coefficients $k_m(\nu )$ given in the Appendix A
(see also Appendix B in Ref. \cite{Kotikov:2022sos}), we have obtained a rather compact form for the derivatives of $\tilde{a}^{(i+1)}_{\nu}(Q^2)$.

\section{MA coupling}

There are several ways to get analytical variants of the strong coupling constant $a_s$ (see, for example, \cite{Bakulev:2008td}). Here we follow the
MA approach \cite{ShS,MSS,Sh} as already discussed in Introduction.
For the fractional case,  the MA approach was generalized by Bakulev, Mikhailov and Stefanis (hereinafter referred to as the BMS approach), which is
presented in three famous papers \cite{BMS1,Bakulev:2006ex,Bakulev:2010gm}
(see also a previous paper \cite{Karanikas:2001cs}, the reviews \cite{Bakulev:2008td,Cvetic:2008bn} and the Mathematica package in \cite{Bakulev:2012sm}).

We will first show the LO BMS results, and later we will go beyond LO, following our results for the ordinary  strong coupling constant obtained in the previous section
(see Eq. (\ref{tdmp1N})).

\subsection{LO}

\begin{figure}[!htb]
\centering
\includegraphics[width=0.58\textwidth]{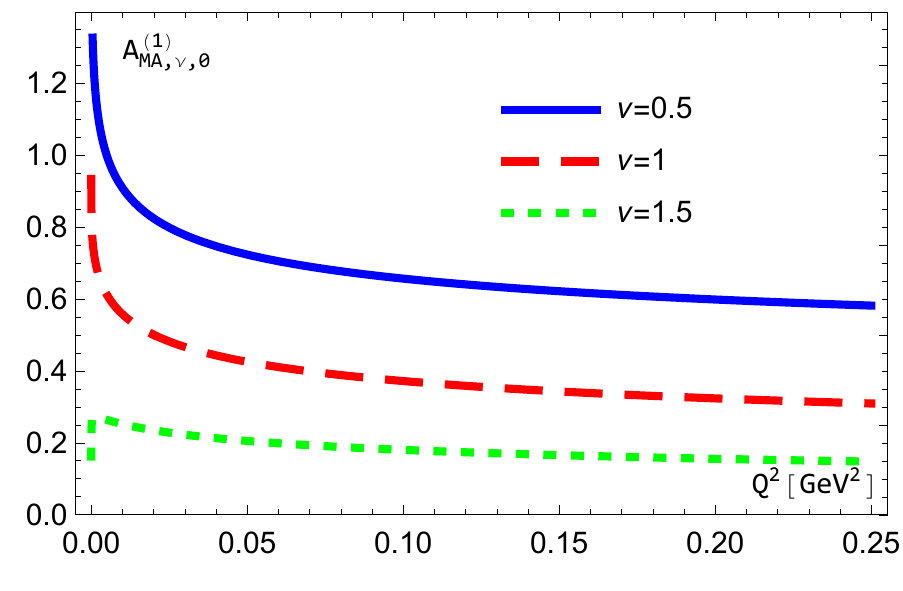}
    \caption{\label{fig:A1}
      The results for $A^{(1)}_{\rm MA,\nu,0}(Q^2)$ with $\nu=0.5,1,1.5$.
    }
\end{figure}

The LO MA coupling $A^{(1)}_{{\rm MA},\nu}$
has the following form  \cite{BMS1}
\be
A^{(1)}_{{\rm MA},\nu,0}(Q^2) = {\left( a^{(1)}_{\nu,0}(Q^2)\right)}^{\nu} - \frac{{\rm Li}_{1-\nu}(z_0)}{\Gamma(\nu)}=
\frac{1}{L_0^{\nu}}- \frac{{\rm Li}_{1-\nu}(z_0)}{\Gamma(\nu)} \equiv \frac{1}{L_0^{\nu}}-\Delta^{(1)}_{\nu,0}\,,
\label{tAMAnu}
\ee
where
\be
   {\rm Li}_{\nu}(z)=\sum_{m=1}^{\infty} \, \frac{z^m}{m^{\nu}}=  \frac{z}{\Gamma(\nu)} \int_0^{\infty} 
\frac{ dt \; t^{\nu -1} }{(e^t - z)}
   \label{Linu}
\ee
is the Polylogarithm. For the cases $\nu=0.5,1,1.5$, $A^{(1)}_{{\rm MA},\nu,0}(Q^2)$ is shown in Fig. \ref{fig:A1}.
\footnote{As we showed in Section 2, the value of the $\Lambda$ parameter is determined by fitting the experimental data. To obtain its values within the framework of
  analytical QCD, it is necessary to approximate the experimental data for various processes using, for example, the formulas obtained here, which simplify the form of
  higher-order terms. However, this requires additional special studies. Therefore, in this article, we use the value of $\Lambda_{f=3}$ obtained in the framework of
  traditional perturbative QCD.}
It is clearly seen that $A^{(1)}_{{\rm MA},\nu,0}(Q^2\to 0)$ agree with its asymptotic values:
\be
A^{(1)}_{{\rm MA},\nu,0}(Q^2= 0) = \left\{
\begin{array}{c}
0 ~~\mbox{when}~~ \nu >1, \\
1 ~~\mbox{when}~~ \nu =1, \\
\infty ~~\mbox{when}~~ \nu <1, 
\end{array}
\right.
   \label{AQ=0}
\ee
obtained in Ref. \cite{Ayala:2018ifo}.

For $\nu=1$ we recover the famous Shirkov-Solovtsov result  \cite{ShS,Sh}:
\be
A^{(1)}_{\rm MA,0}(Q^2) \equiv A^{(1)}_{\rm MA,\nu=1,0}(Q^2) =  a^{(1)}_{s,0}(Q^2) - \frac{z_0}{1-z_0}=\frac{1}{L_0}- \frac{z_0}{1-z_0}\, ,
\label{tAM1}
\ee
since
\be
   {\rm Li}_{0}(z)= \frac{z}{1-z} \, .
\label{Li0.1}
\ee
Note that the result (\ref{tAM1}) can be taken directly for the integral form (\ref{disp_MA_LO}), as it was in Ref. \cite{ShS,Sh}.

\subsection{Beyond LO}

Following Eq. (\ref{tAMAnu}) for the LO analytic coupling constant, 
we consider the difference between the derivatives of the  ordinary and MA coupling constants, shown in Eq.(\ref{tan+1}) and in
\be
\tilde{A}_{{\rm MA},n+1}(Q^2)=\frac{(-1)^n}{
  n!} \, \frac{d^n A_{\rm MA}(Q^2)}{(dL)^n} \, ,
\label{tanMA+1}
\ee
respectively.

By analogy with ordinary coupling constant,
using the results (\ref{tdmp1N}) for the differences of fractional derivatives of conventional and MA  coupling constants
\be
\tilde{\Delta}^{(i+1)}_{\nu,i} \equiv \tilde{a}^{(i+1)}_{\nu,i} - \tilde{A}^{(i+1)}_{{\rm MA},\nu,i}
\label{tDeltaMAnu}
\ee
we have the following results
\be
\tilde{\Delta}^{(i+1)}_{\nu,i}
=\tilde{\Delta}^{(1)}_{\nu,i}
+\sum_{m=1}^i C_m^{\nu+m} \, \hat{R}_m \left( \frac{{\rm Li}_{-\nu-m+1}(z_i)}{\Gamma(\nu+m)}\right)
\, ,
\label{tAMAnu.1}
\ee
where the operators $\hat{R}_i$ $(i=1,2)$ are shown above in Eq. (\ref{hR_i}).

Thus, in order to obtain (\ref{tAMAnu.1}), we propose that the form (\ref{tdmp1N}) for an ordinary $a_s$ coupling constant at high orders is just as applicable
to the case of MA coupling constant.

After some evaluations, we obtain the following expressions without operators
\be
\tilde{\Delta}^{(i+1)}_{\nu,i}
=\tilde{\Delta}^{(1)}_{\nu,i}
+\sum_{m=1}^i C_m^{\nu+m} \, \overline{R}_m(z_i) \left( \frac{{\rm Li}_{-\nu-m+1}(z_i)}{\Gamma(\nu+m)}\right)
\, ,
\label{tAMAnu.2}
\ee
where
\bea
&&\overline{R}_1(z)=b_1\Bigl[\gamma_{\rm E}-1+{\rm M}_{-\nu,1}(z)\Bigr], \nonumber \\
&&\overline{R}_2(z)=b_2 + b_1^2\Bigl[{\rm M}_{-\nu-1,2}(z) + 2(\gamma_{\rm E}-1){\rm M}_{-\nu-1,1}(z) +  (\gamma_{\rm E}-1)^2-
  \zeta_2\Bigr]
\label{oRi}
\eea
and
\be
   {\rm Li}_{\nu,k}(z)=(-1)^k\,\frac{d^k}{(d\nu)^k}  \,{\rm Li}_{\nu}(z) =
   \sum_{m=1}^{\infty} \, \frac{z^m\ln^k m}{m^{\nu}},~~{\rm M}_{\nu,k}(z)=\frac{{\rm Li}_{\nu,k}(z)}{{\rm Li}_{\nu}(z)} \, .
   \label{Mnuk}
\ee
As we can see, the $\Psi(\nu)$-function and its derivatives have completely canceled out.

So, we have for MA analytic coupling constants $\tilde{A}^{(i+1)}_{{\rm MA},\nu}$ the following expressions:
\be
\tilde{A}^{(i+1)}_{{\rm MA},\nu,i}(Q^2) = \tilde{A}^{(1)}_{{\rm MA},\nu,i}(Q^2) + \sum_{m=1}^{i}  \, C^{\nu+m}_m \tilde{\delta}^{(m+1)}_{{\rm ma},\nu,i}(Q^2) \,
\label{tAiman}
\ee
where
\bea
&&\tilde{A}^{(1)}_{{\rm MA},\nu,i}(Q^2) = \tilde{a}^{(1)}_{\nu,i}(Q^2) -  \frac{{\rm Li}_{1-\nu}(z_i)}{\Gamma(\nu)},
~~\nonumber \\
&&\tilde{\delta}^{(m+1)}_{{\rm MA},\nu,i}(Q^2)= \tilde{\delta}^{(m+1)}_{\nu,i}(Q^2) -  \overline{R}_m(z_i)   \, \frac{{\rm Li}_{-\nu+1-m}(z_i)}{\Gamma(\nu+m)}
\label{tdAman}
\eea
and $\tilde{\delta}^{(k+1)}_{\nu,m}(Q^2)$
are given in Eq. (\ref{tdmp1N}).

\subsection{MA coupling itself}

We would like to note that the results (\ref{tAMAnu.1}) can be rewritten in the following way
\be
\tilde{\Delta}^{(i+1)}_{\nu,i}=\tilde{\Delta}^{(1)}_{\nu,i}
+\sum_{m=1}^i \, \frac{P_{m,\nu}(z_i)}{m! \Gamma(\nu)}
\,,~~ P_{m,\nu}(z_i)= \overline{R}_m(z_i) \, {\rm Li}_{-\nu-m+1}(z_i)
\label{tAMAnu.2a}
\ee
where
\bea
&&P_{1,\nu}(z)=b_1\Bigl[\Bigl(\gamma_{\rm E}-1\Bigr){\rm Li}_{-\nu}(z)+{\rm Li}_{-\nu,1}(z)\Bigr], \nonumber \\
&&P_{2,\nu}(z)=b_2 \,{\rm Li}_{-\nu-1}(z) + b_1^2\Bigl[{\rm Li}_{-\nu-1,2}(z) + 2(\gamma_{\rm E}-1){\rm Li}_{-\nu-1,1}(z) \nonumber \\
&&+  \Bigl((\gamma_{\rm E}-1)^2-
  \zeta_2\Bigr) \, {\rm Li}_{-\nu-1}(z) \Bigr]\,.
\label{Pkz}
\eea

For the case $\nu=1$, at LO we have Eq. (\ref{tAM1}) and above LO we can apply above results (\ref{tAiman}) - (\ref{tdAman})
to the case $\nu=1$:
\be
A^{(i+1)}_{{\rm MA},i}(Q^2)\equiv \tilde{A}^{(i+1)}_{{\rm MA},\nu=1,i}(Q^2) = A^{(1)}_{{\rm MA},i}(Q^2) + \sum_{m=1}^{i}  \, \tilde{\delta}^{(m+1)}_{{\rm MA},1,i}(Q^2) \,
\label{tAiman.1}
\ee
where (according to (\ref{tAM1}))
\bea
&&A^{(1)}_{{\rm MA},i}(Q^2) = \tilde{a}^{(1)}_{\nu=1,i}(Q^2) -  {\rm Li}_{0}(z_i)= a^{(1)}_{s,i}(Q^2) -  {\rm Li}_{0}(z_i),
~~\nonumber \\
&&\tilde{\delta}^{(m+1)}_{{\rm MA},1,i}(Q^2)= \tilde{\delta}^{(m+1)}_{1,i}(Q^2) -  \overline{R}_m(z_i)   \, \frac{{\rm Li}_{-m}(z_i)}{m!}
= \tilde{\delta}^{(m+1)}_{1,i}(Q^2) - \frac{P_{m,1}(z_i)}{m!} \, ,
\label{tdAmanA}
\eea
with $P_{m,1}(z_i)$ are given in Eq. (\ref{tAMAnu.2a})
at $\nu=1$, ${\rm Li}_{0}(z)$ in (\ref{Li0.1})
and also
\bea
    {\rm Li}_{-1}(z)= \frac{z}{(1-z)^2},~~{\rm Li}_{-2}(z)= \frac{z(1+z)}{(1-z)^3}
    \, .
\label{Lii.1}
\eea
Eqs. (\ref{tdAmanA}) and (\ref{Lii.1}) can be used for applications beyond LO within the MA QCD.

\subsection{Results}

\begin{figure}[!htb]
\centering
\includegraphics[width=0.58\textwidth]{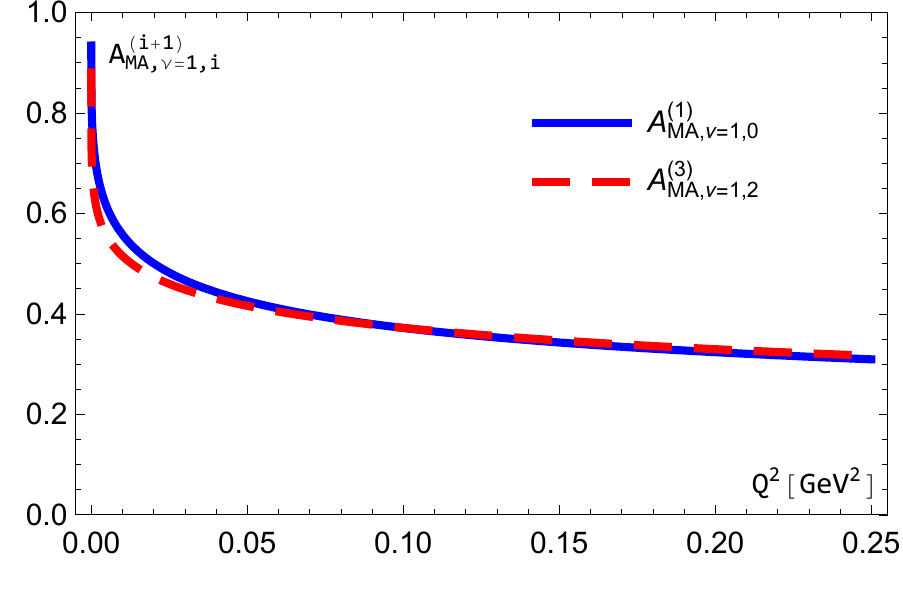}
    \caption{\label{fig:A123}
      The results for $A^{(i+1)}_{\rm MA,\nu=1,i}(Q^2)$ with $i=0,2$.}
\end{figure}

\begin{figure}[!htb]
\centering
\includegraphics[width=0.58\textwidth]{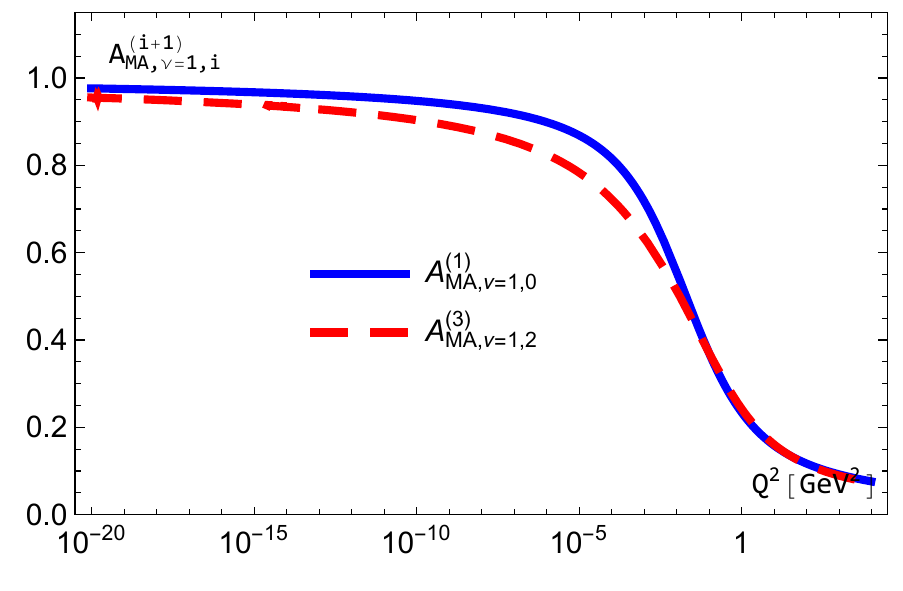}
    \caption{\label{fig:A123LOG}
Same as in Fig. (\ref{fig:A123}) but with the logarithmic scale.
    }
\end{figure}

\begin{figure}[!htb]
\centering
\includegraphics[width=0.58\textwidth]{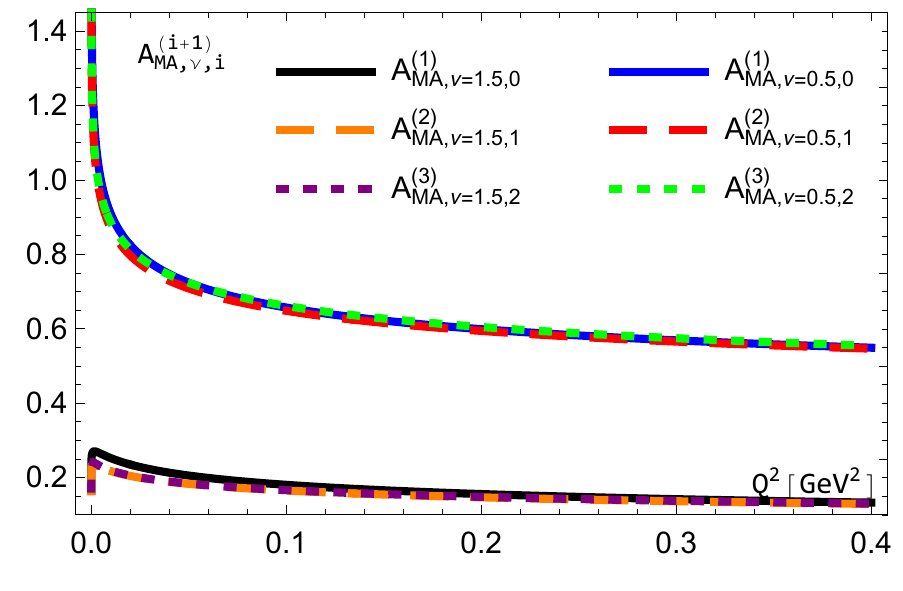}
    \caption{\label{fig:fracALL}
      The results for $A^{(i+1)}_{\rm MA,\nu=0.5,i}(Q^2)$ and $A^{(i+1)}_{\rm MA,\nu=1.5,i}(Q^2)$ with $i=0,1,2$.
    }
\end{figure}

\begin{figure}[!htb]
\centering
\includegraphics[width=0.58\textwidth]{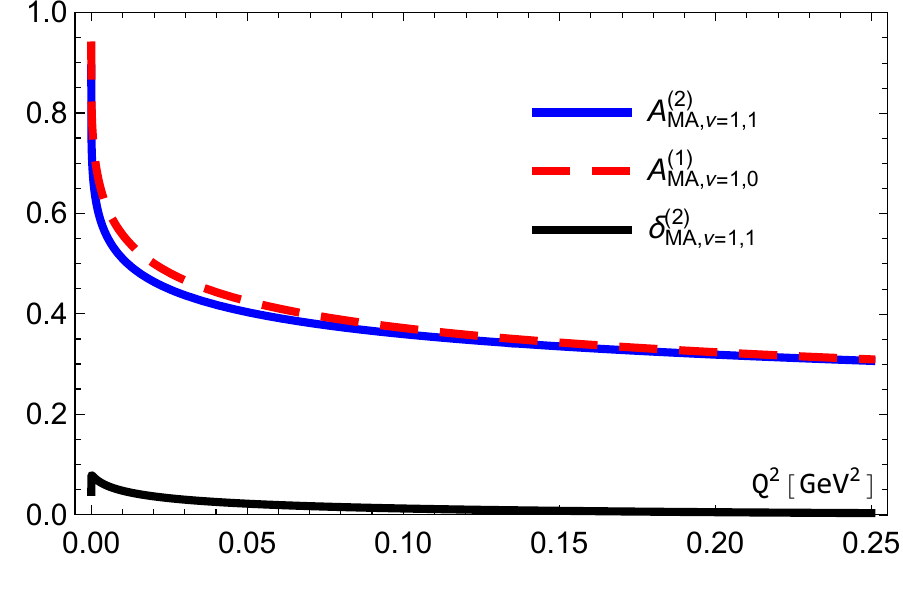}
    \caption{\label{fig:Adelta2}
      The results for $A^{(1)}_{\rm MA,\nu=1,0}(Q^2)$, $A^{(2)}_{\rm MA,\nu=1,1}(Q^2)$ and $\delta^{(2)}_{\rm MA,\nu=1,1}(Q^2)$.
    }
\end{figure}

\begin{figure}[!htb]
\centering
\includegraphics[width=0.58\textwidth]{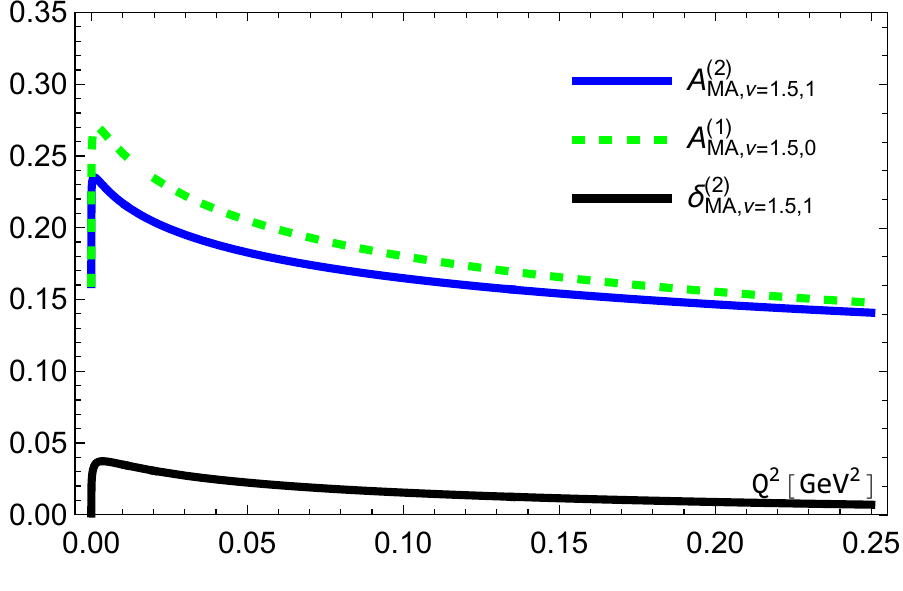}
    \caption{\label{fig:Afrac2}
      The results for $A^{(1)}_{\rm MA,\nu=1.5,0}(Q^2)$, $A^{(2)}_{\rm MA,\nu=1.5,1}(Q^2)$ and $\delta^{(2)}_{\rm MA,\nu=1.5,1}(Q^2)$.
    }
\end{figure}

\begin{figure}[!htb]
\centering
\includegraphics[width=0.58\textwidth]{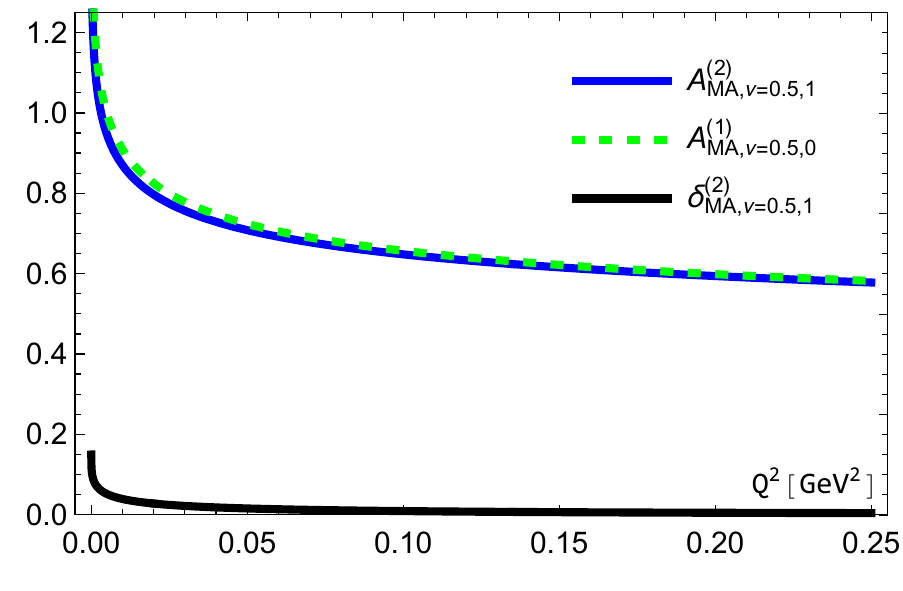}
    \caption{\label{fig:Afrac3}
      The results for $A^{(1)}_{\rm MA,\nu=0.5,0}(Q^2)$, $A^{(2)}_{\rm MA,\nu=0.5,1}(Q^2)$ and $\delta^{(2)}_{\rm MA,\nu=0.5,1}(Q^2)$.
    }
\end{figure}

\begin{figure}[!htb]
\centering
\includegraphics[width=0.58\textwidth]{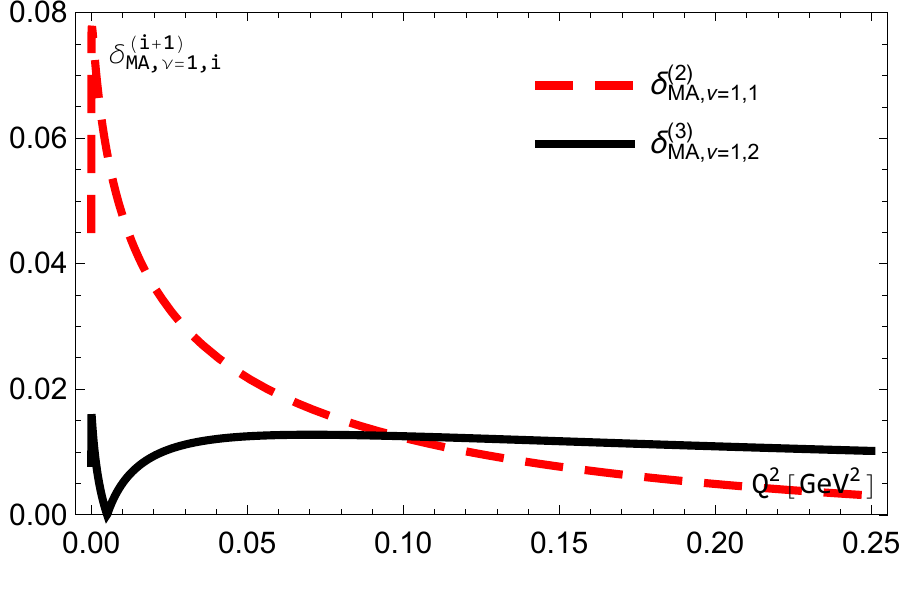}
    \caption{\label{fig:deltaALL}
      The results for $\delta^{(i+1)}_{\rm MA,\nu=1,i}(Q^2)$ with $i=1,2$.}
\end{figure}

\begin{figure}[!htb]
\centering
\includegraphics[width=0.58\textwidth]{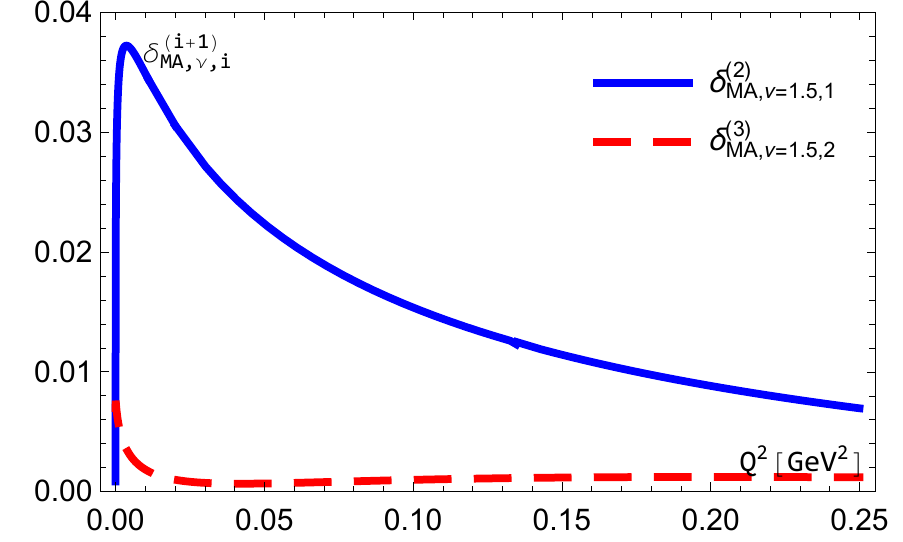}
    \caption{\label{fig:deltafrac32sec}
      The results for $\delta^{(i+1)}_{\rm MA,\nu=1.5,i}(Q^2)$ with $i=1,2$.}
\end{figure}

As can be seen from Figs. \ref{fig:A123} and \ref{fig:A123LOG} differences between $A^{(i+1)}_{\rm MA,\nu=1,i}(Q^2)$ with $i= 0.2$ are small and
have nonzero values only for $Q^2=\Lambda_i^2$. A similar situation exists in the cases $\nu=0.5$ and $\nu=1.5$ (see fig. \ref{fig:fracALL}).

On Figs. \ref{fig:Adelta2}, \ref{fig:Afrac2} and \ref{fig:Afrac3} one can see that the difference $\delta^{(2)}_{\rm MA,\nu,1}(Q^2)$ is significantly less than
the coupling constants themselves. This is shown for three different values of $\nu$: $0.5, 1, 1.5$.
From Figs. \ref{fig:fracALL}, \ref{fig:Afrac2} and \ref{fig:Afrac3} it is clear that for $Q^2\to 0$ the asymptotic behavior of $A^{(1)}_{\rm MA,\nu, 0}(Q^2)$ and
$A^{(2)}_{\rm MA,\nu,1}(Q^2)$ coincide (and are equal to those considered in (\ref{AQ=0})) i.e. differences $\delta^{(1)}_{\rm MA,\nu=1,1}(Q^2\to 0)$ and
$\delta ^{(2)}_{\rm MA,\nu =1,2}(Q^2\to 0)$ tend to zero values. Fig. \ref{fig:deltaALL} shows the difference
$\delta^{(3)}_{\rm MA,\nu=1,i}(Q^2)$ $(i\geq 2)$, which are substantially less than $\delta^{(1)} _{\rm MA,\nu=1,1}(Q^2)$. From Fig. \ref{fig:deltafrac32sec} one can see
a similar property for the $\nu=1.5$ case.

Thus, we can conclude that, in contrast to the case of a conventional coupling constant, considered in Fig. 1, the $1/L$-expansions of the MA coupland is a very good
approximation for any values of $Q^2$.
Moreover, the differences between $A^{(i+1)}_{\rm MA,\nu=1,i}(Q^2)$ and $A^{(1)}_{\rm MA,\ nu=1,0}(Q^2)$ are small.
Therefore, the expansions of $A^{(i+1)}_{\rm MA,\nu=1,i}(Q^2)$ $i\geq 1$ in $A^{(1)}_{\rm MA,\nu=1,0}(Q^2)$ made in Refs. \cite{BMS1,Bakulev:2006ex,Bakulev:2010gm}, are
very good approximations.
Also the approximation
\be
A^{(i+1)}_{\rm MA,\nu=1,i}(Q^2)=A^{(1)}_{\rm MA,\nu=1,0}(k_iQ^2),~~(i=1,2)\,,
\label{Appro}
\ee
introduced in \cite{Pasechnik:2009yc,Khandramai:2011zd} and used in \cite{Kotikov:2010bm} is very convenient, too.
Indeed, since the corrections $\delta^{(i+1)}_{\rm MA,\nu=1,i}(Q^2)$ are very small, then for example from Eq. (\ref{tdAmanA}) one can see that the MA coupling constants
$A^{(i+1)}_{\rm MA,\nu=1,i}(Q^2)$ are very similar to the LO ones taken with the same $\Lambda_i$.

\section{MA coupling. Another form}

The results (\ref{tAMAnu.2}) for MA  coupling constant are very convenient in the range of large and small values of $Q^2$. For $Q^2 \sim \Lambda_i^2$ both parts,
the standard strong coupling constant and the additional term, have singularities that cancel out in sum. Thus, numerical applications of the results
(\ref{tAMAnu.2}) may not be so easy, requiring, for example, some sub-expansions for each part in the neighborhood of the point $Q^2 =\Lambda_i^2$ .
Therefore, here we propose another form that is very useful for $Q^2 \sim \Lambda_i^2$ and can be used for any value of $Q^2$ as well, except for the
ranges of very large and very small $Q^ 2$ values. As in the previous section, we will first present the LO results, taken from the excellent paper \cite{BMS1},
and then extend them beyond LO.

\subsection{LO}

The LO MA coupling $A^{(1)}_{{\rm MA},\nu}(Q^2)$ \cite{ShS,MSS,Sh}
have also the another form  \cite{BMS1}
\be
A^{(1)}_{{\rm MA},\nu}(Q^2) =
\frac{(-1)}{\Gamma(\nu)} \, \sum_{r=0}^{\infty} \zeta(1-\nu-r) \, \frac{(-L)^r}{r!}~~~ (L<2\pi),
\label{tAMAnuNew}
\ee
where Euler functions $\zeta(\nu)$ are
\be
\zeta(\nu)
     =\sum_{m=1}^{\infty} \, \frac{1}{m^{\nu}}={\rm Li}_{\nu}(z=1)
   \label{ze_nu}
\ee

The result (\ref{tAMAnuNew}) was obtained in Ref. \cite{BMS1} using properties of the Lerch function, which can be considered as a
generalization of Polylogarithms (\ref{Linu}). The form (\ref{tAMAnuNew}) is very convenient for low values of $L$, i.e. in $Q^2 \sim \Lambda^2$.

Moreover, it is possible to use the relation between $\zeta(1-\nu-r)$ and $\zeta(\nu+r)$ functions
\be
\zeta(1-\nu-r)= \frac{2\Gamma(\nu+r)}{(2\pi)^{\nu+r}}\, Sin\left[\frac{\pi}{2}(1-\nu-r)\right] \, \zeta(\nu+r)
  \label{ze_nu.1}
\ee

For $\nu=1$ we have
\be
A^{(1)}_{\rm MA}(L) = - \, \sum_{r=0}^{\infty} \zeta(-r) \, \frac{(-L)^r}{r!},~~ \zeta(-r) = (-1)^{r} \, \frac{B_{r+1}}{r+1}
\label{tAMA1New}
\ee
where $B_{r+1}$ are Bernoulli numbers.
Using their properties, we have for even $r=2m$ and for odd  $r=1+2l$ values
\be
\zeta(-2m) = -\frac{\delta^0_m}{2},~~\zeta(-(1+2l))= - \frac{B_{2(l+1)}}{2(l+1)} \,,
\label{ze_r1}
\ee
where $\delta^0_m$ is the Kronecker symbol.

Thus, for $A^{(1)}_{\rm MA}(Q^2)$  we have 
\be
A^{(1)}_{\rm MA}(Q^2) = \frac{1}{2} \, \left(1
+ \, \sum_{l=0}^{\infty} \frac{B_{2(l+1)}}{l+1} \, \frac{(-L)^{2l+1}}{(2l+1)!}\right)=\frac{1}{2} \, \left(1
+ \, \sum_{s=1}^{\infty} \frac{B_{2s}}{s} \, \frac{(-L)^{2s-1}}{(2s-1)!}\right) \, ,
\label{tAMA1New.1}
\ee
with $s=l+1$.

\subsection{Beyond LO}

Now we consider the derivatives of MA coupling constant, i.e. $\tilde{A}^{(1)}_{{\rm MA},\nu}$, shown
in Eq. (\ref{tAiman}), i.e.
\be
\tilde{A}^{(i+1)}_{{\rm MA},\nu,i}(Q^2) = \tilde{A}^{(1)}_{{\rm MA},\nu,i}(Q^2) + \sum_{m=1}^{i}  \, C^{\nu+m}_m \tilde{\delta}^{(m+1)}_{{\rm MA},\nu,i}(Q^2) \,
\label{tAimanNew}
\ee
where $\tilde{A}^{(1)}_{{\rm MA},\nu,i}=A^{(1)}_{{\rm MA},\nu}$ is given above in (\ref{tAMAnuNew}) with $L \to L_{i}$ and
\be
\tilde{\delta}^{(m+1)}_{{\rm MA},\nu,i}(Q^2)= \hat{R}_m   \, A^{(1)}_{{\rm MA},\nu+m,i} \, ,
\label{tdAmanNew}
\ee
where operators $\hat{R}_m$ are given above in (\ref{hR_i}).

After some calculations we have
\be
\tilde{\delta}^{(m+1)}_{{\rm MA},\nu,k}(Q^2)=
\frac{(-1)}{\Gamma(\nu+m)} \, \sum_{r=0}^{\infty} \tilde{R}_m(\nu+r) \, \frac{(-L_k)^r}{r!}
\label{tdAmanNew}
\ee
where, in an agreement with (\ref{oRi}),
\bea
&&\tilde{R}_1(\nu+r)=b_1\Bigl[(\gamma_{\rm E}-1)\zeta(-\nu-r)+\zeta_1(-\nu-r)\Bigr], \nonumber \\
&&\tilde{R}_2(\nu+r)=b_2\zeta(-\nu-r-1) + b_1^2\Bigl[\zeta_2(-\nu-r-1) + 2(\gamma_{\rm E}-1)\zeta_1(-\nu-r-1) \nonumber \\
  &&\hspace{0.5cm} +  \bigl[(\gamma_{\rm E}-1)^2-
    \zeta_2\bigr]\zeta(-\nu-r-1)\Bigr]
\label{tRi}
\eea
and
\be
\zeta_k(\nu)=  {\rm Li}_{\nu,k}(z=1)=\sum_{m=1}^{\infty} \, \frac{\ln^k m}{m^{\nu}}
\, .
   \label{zetaknu}
\ee

Strictly speaking, the functions $\zeta_n(-m-\nu-r -k)$ in (\ref{tRi})
are not so good defined at large $r$ values and we can replace them by $\zeta_n(m+\nu+r+k)$
using the result (\ref{ze_nu.1}). The results are presented in Appendix C.

\subsection{MA coupling itself}

For the case $\nu=1$ we immediately have
\bea
&&A^{(i+1)}_{\rm MA,i}(Q^2) = A^{(1)}_{\rm MA,i}(Q^2) + \sum_{m=1}^{i}  \, \tilde{\delta}^{(m+1)}_{\rm MA,\nu=1,i}(Q^2) \, ,\label{tdAmanNew.1a}\\
&&\delta^{(m+1)}_{\rm MA,i}(L) \equiv \tilde{\delta}^{(m+1)}_{\rm MA,\nu=1,i}(Q^2) = 
\frac{(-1)}{m!} \, \sum_{r=0}^{\infty} \tilde{R}_m(1+r) \, \frac{(-L_i)^r}{r!} \, ,
\label{tdAmanNew.1}
\eea
where $A^{(1)}_{\rm MA,i}(Q^2)$ is given above in (\ref{tAMA1New}), with the replacement $L \to L_i$,  and the coefficients
$\tilde{R}_m(1+r)$ can be found in (\ref{tRi}) when $\nu=1$.

The results (\ref{tdAmanNew.1}) can be expressed in terms of the functions $\zeta_n(m+\nu+r+k)$. Using the results shown in Appendix C
and taking separately the even part ($r=2m$) and the odd part ($r=2s-1$) (see Eq. (\ref{tdAmanNew})), we have 
\bea
\z \delta^{(2)}_{{\rm MA},k}(Q^2)= \frac{2}{(2\pi)^2} \, \Biggl[\sum_{m=0}^{\infty} (2m+1)(-1)^m Q_{1a}(2m+2)\hat{L}_k^{2m} -\pi \sum_{s=1}^{\infty} s(-1)^s Q_{1b}(2s+1)\hat{L}_k^{2s-1}
  \Biggr], \nonumber \\
\z \delta^{(3)}_{{\rm MA},k}(Q^2)= -\frac{1}{(2\pi)^3} \, \Biggl[\pi \sum_{m=0}^{\infty} (2m+1)(m+1)(-1)^m Q_{2b}(2m+3)\hat{L}_k^{2m} \nonumber \\&&\hspace{1cm}
  +2 \sum_{s=1}^{\infty} s(2s+1)(-1)^s Q_{2a}(2s+2)\hat{L}_k^{2s-1}
  \Biggr],~~ \hat{L}_k=\frac{L_k}{2\pi}\,,
\label{tdAmanNew1}
\eea
where
the function $Q_{ma}$ and $Q_{mb}$ are given in Appendix B.

At the point $L_k=0$, i.e. for $Q^2=\Lambda_k^2$, we get
\bea
&&A^{(1)}_{\rm MA}= \frac{1}{2},~~~
\delta^{(2)}_{s}= \frac{2}{(2\pi)^2} \, Q_{1a}(2)= -\frac{b_1}{2\pi^2} \, \Bigl(\zeta_1(2)+l\zeta(2)\Bigr),\nonumber \\
&&\delta^{(3)}_{s}= -\frac{\pi}{(2\pi)^3} \, Q_{2b}(3)=  \frac{b_1^2}{4\pi^2} \, \Bigl(\zeta_1(3)+(2l-1)\zeta(3)\Bigr),
\label{dAmanNew2}
\eea
where $\zeta_k(\nu)$ are given in Eq. (\ref{zetaknu}) and $l=\ln(2\pi)$.

\section{Integral representations for MA coupling }

As already discussed in Introduction, the MA coupling constant $A^{(1)}_{\rm MA}(Q^2)$ is constructed as follows:
the LO spectral function is taken directly from perturbation theory, and the MA coupling constant $A^{(1)}_{\rm MA}(Q^2)$ is obtained from the dispersion integral
using the correct integration contour. So, at LO, the MA coupling constant $A^{(1)}_{\rm MA}(Q^2)$ obeys the equation (\ref{disp_MA_LO}) presented in Introduction.

For the $\nu$-derivative of $A^{(1)}_{\rm MA}(Q^2)$, i.e. $\tilde{A}^{(1)}_{\rm MA,\nu}(Q^2)$, there is the following
equation \cite{GCAK}:
\be
\tilde{A}^{(1)}_{\rm MA,\nu}(Q^2)=
\frac{(-1)}{
  \Gamma(\nu)}
\int_{0}^{\infty} \ \frac{d s}{s} r^{(1)}_{\rm pt}(s)
    {\rm Li}_{1-\nu} (-sz)\,,
\label{disptAnuz} 
\ee
where ${\rm Li}_{1-\nu} (-sz)$ is the Polylogarithm presented in Eq. (\ref{Linu}).

Beyond LO, Eq. (\ref{disptAnuz}) can be extended
in two different ways, which will be shown in following subsections.

\subsection{Modification of spectral functions}

The first possibility to extend the result (\ref{disptAnuz}) beyond LO is related to the modification of the spectral function:
\be
\tilde{A}^{(i+1)}_{{\rm MA},\nu,k}(Q^2) =
\frac{(-1)}{
  \Gamma(\nu)}
\int_{0}^{\infty} \ \frac{d s}{s} r^{(i+1)}_{\rm pt}(s)
    {\rm Li}_{1-\nu} (-sz_k)\,,
    \label{disptAnuz.ma}
    \ee
    i.e. it is similar to (\ref{disptAnuz}) with the replacement the LO spectral function $r^{(1)}_{\rm pt}(s)$ by $i+1$-order one $r^{(i+1)}_{\rm pt}(s)$,
    which have the following form
\be
r^{(i+1)}_{\rm pt}(s)=r^{(1)}_{\rm pt}(s)+ \sum_{m=1}^{i} \delta^{(m+1)}_{\rm r}(s) \, 
\label{rima.1}
\ee
and
(see \cite{NeSi,Nesterenko:2017wpb})
\be
y=\ln s,~r^{(1)}_{\rm pt}(y)=\frac{1}{y^2+\pi^2} \,,~
\delta^{(2)}_{\rm r}(y)=-\frac{b_1}{(y^2+\pi^2)^2} \, \Bigl[2y f_1(y) +(\pi^2-y^2) f_2(y)\Bigr] \,,
\label{dr5ma}
\ee
with
\be
f_1(y)=\frac{1}{2} \, \ln\bigl(y^2+\pi^2\bigr),~~f_2(y)=\frac{1}{2} -\frac{1}{\pi}\, arctan\left(\frac{y}{\pi}\right) \, .
 \label{f12}
\ee

  For the MA coupling constant itself, we have
   \be
   A^{(i+1)}_{\rm MA,k}(Q^2) \equiv \tilde{A}^{(i+1)}_{{\rm MA},\nu=1,k}(Q^2)    =
   \int_0^{+\infty} \,  
\frac{ d s \, r^{(i+1)}_{\rm pt}(s)}{(s + t_k)}\,.
\label{dispA.ma} 
\ee

Numerical evaluations af the integrals in (\ref{dispA.ma}) can be done following to discussions in Section 4 in Ref. \cite{NeSi}.

\subsection{Modification of Polylogaritms}

Beyond LO, the results (\ref{disptAnuz}) can also be expanded with the $\hat{R}_m$ operators shown in (\ref{hR_i}), and this results in the following result:
\be
\tilde{A}^{(i+1)}_{{\rm MA},\nu,i}(Q^2) =
\int_{0}^{\infty}  \frac{d s}{s} r^{(1)}_{\rm pt}(s)
\tilde{\Delta}^{(i+1)}_{\nu,i}\,,
    \label{disptAnuz.ma1}
    \ee
where the results for $\tilde{\Delta}^{(i+1)}_{\nu,i}$ can be found in Eqs. (\ref{tAMAnu.1}), (\ref{tAMAnu.2}), (\ref{oRi}) and also in Eqs. (\ref{tAMAnu.2a})-(\ref{Pkz}).

For MA coupling constant itself, we have
\be
A^{(i+1)}_{\rm ma,i}(Q^2) \equiv \tilde{A}^{(i+1)}_{{\rm ma},\nu=1,i}(Q^2) =
\int_0^{+\infty} \,  
 \frac{d s}{s} r^{(1)}_{\rm pt}(s)
\tilde{\Delta}^{(i+1)}_{\nu=1,i}\, ,
\label{dispA.ma2} 
\ee
where the results for $\tilde{\Delta}^{(i+1)}_{\nu=1,i}$ are given in Eq. (\ref{tAMAnu.2a}) with $\nu=1$, i.e. 
\be
\tilde{\Delta}^{(i+1)}_{\nu=1,i}=\tilde{\Delta}^{(1)}_{1,i} +\sum_{m=1}^i \, \frac{P_{m,1}(z_i)}{m!}=\Delta^{(1)}_{1,i} +\sum_{m=1}^i \, \frac{P_{m,1}(z_i)}{m!}\,,
\label{tAMAnu.2aN}
\ee
where $\Delta^{(1)}_{1,i}={\rm Li}_0(z_i)$ and  $P_{m,1}(z_i)$ are given in Eq. (\ref{tAMAnu.2a})
at $\nu=1$.

\subsection{Results}

\begin{figure}[!htb]
\centering
\includegraphics[width=0.58\textwidth]{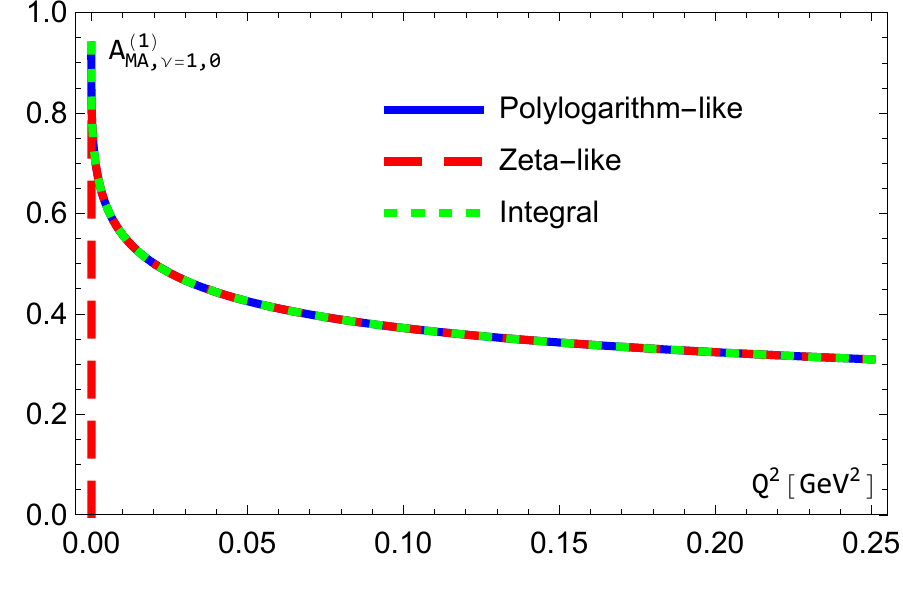}
    \caption{\label{fig:A1Leg}
      The results for $A^{(1)}_{\rm MA,\nu=1,0}(Q^2)$ with
      $\Lambda_0(f=3)$ done in Eq. (\ref{Lambdas}). The Polylogarithm-like, zeta-like and integral (\ref{disptAnuz.ma}) forms
      have been used.}
\end{figure}

\begin{figure}[!htb]
\centering
\includegraphics[width=0.58\textwidth]{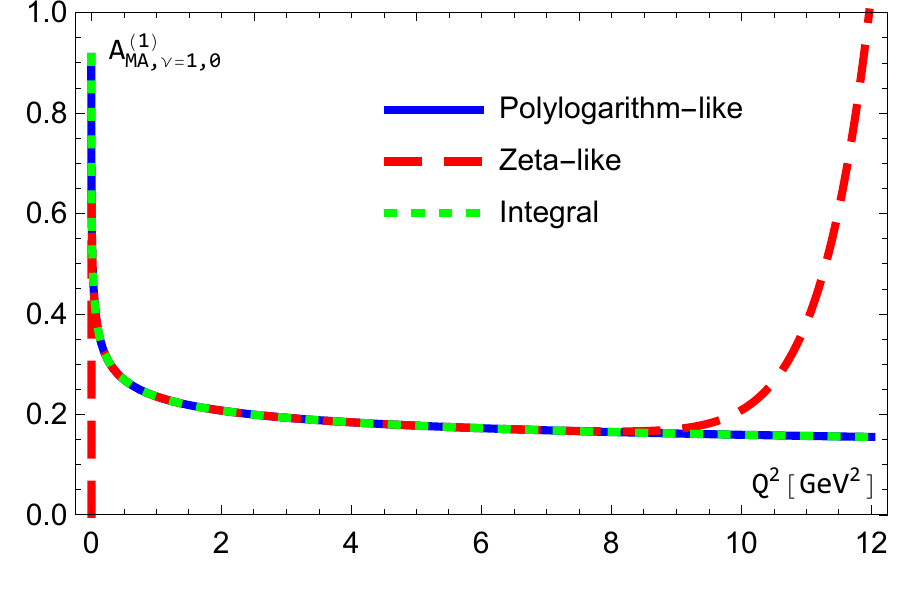}
    \caption{\label{fig:A1wide}
Same as in Fig.(\ref{fig:A1wide}) but for larger $Q^2$ values.}
\end{figure}

\begin{figure}[!htb]
\centering
\includegraphics[width=0.58\textwidth]{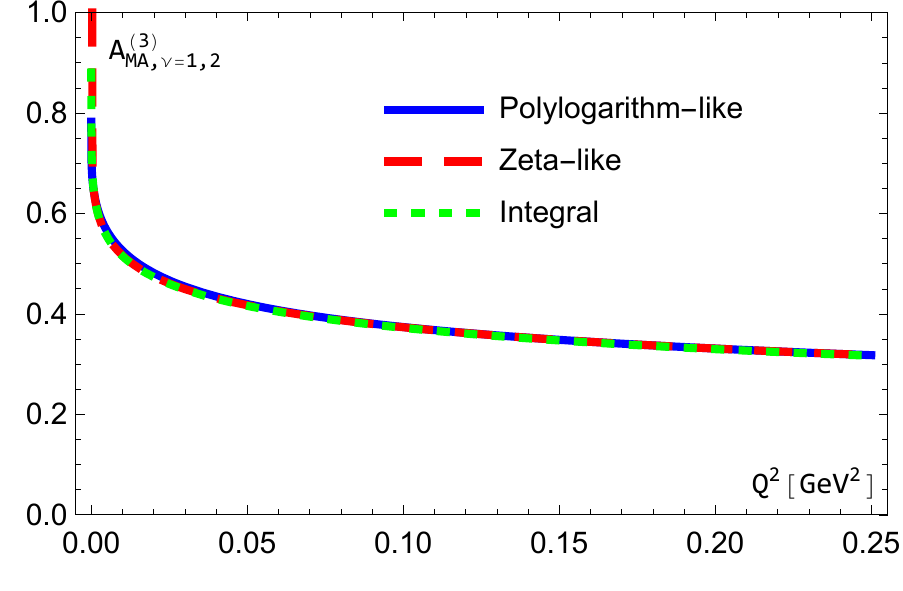}
    \caption{\label{fig:All3}
      The results for $A^{(3)}_{\rm MA,\nu=1,2}(Q^2)$ with
      $\Lambda_2(f=3)$ done in Eq. (\ref{Lambdas}). The Polylogarithm-like, zeta-like and integral forms
      have been used.}
\end{figure}

From Figs. \ref{fig:A1Leg}, \ref{fig:A1wide} and \ref{fig:All3} it can be seen that the results (\ref{tAiman.1}) (we call this "polylogarithm-like")
and the results (\ref{disptAnuz.ma}) (we call them "integral forms") are exactly the same. The results (\ref{tdAmanNew.1a}), (\ref{tdAmanNew.1})
do not apply well for $Q^2 \to 0$ and $Q^2 \to \infty$, as they should, because we actually use only a finite number of terms $(r\leq 100)$
in the sum on the right side (\ref{tdAmanNew.1a}).
However, the results of (\ref{tdAmanNew.1a})-(\ref{tdAmanNew1}) are very good for $Q^2$ intermediate values: $0.1$ GeV$^2< Q^2< 10$ GeV$^2$.

As an application of the MA coupling constant, we study the Bjorken sum rule.
We follow previous research in Refs. 
\cite{Pasechnik:2009yc,Khandramai:2011zd,Pasechnik:2008th,Ayala:2017uzx,Ayala:2018ulm,Chen:2006tw,Chen:2005tda,Kotikov:2012eq}
(see also the Charter IV.8 in Ref. \cite{Enterria}).

\section{Bjorken sum rule}

The polarized Bjorken sum rule is defined as the difference between the proton and neutron polarized structure functions, integrated over
the entire interval $x$
\be
\Gamma_1^{p-n}(Q^2)=\int_0^1 \, dx\, \bigl[g_1^{p}(x,Q^2)-g_1^{n}(x,Q^2)\bigr].
\label{Gpn} 
\ee

Theoretically, the quantity can be written in the Operator Produxt Expansion
form
\be
\Gamma_1^{p-n}(Q^2)=\left|\frac{g_A}{g_V}\right| \, \frac{1}{6} \, \bigl(1-D_{\rm BS}(Q^2)\bigr) + \sum_{i=2}^{\infty} \frac{\mu_{2i}(Q^2)}{Q^{2i-2}} \, ,
\label{Gpn.OPE} 
\ee
where $|g_A/g_V|$=1.2723 $\pm$ 0.0023 is the ratio of the nucleon axial charge, $(1-D_{BS}(Q^2))$ is the leading-twist
contribution, and $\mu_{2i}/Q^{2i-2}$ is the higher-twist (HT)
contribution.

The
twist-four term \cite{Shuryak:1981kj} can be expressed at LO
\footnote{For the HT corrections we restricte ourselves by LO approximation.}
as \cite{Chen:2006tw,Chen:2005tda}
(see discussions in Ref.  \cite{Pasechnik:2009yc}):
\be
\mu_{4}(Q^2)=\mu_{4}(Q^2_0) \, {\left[\frac{a_s(Q^2)}{a_s(Q^2_0)}\right]}^{d_4},~~ d_4=\frac{32}{9\beta_0}\, ,
\label{mu4Q2} 
\ee
which is modified in the MA QCD as (see \cite{Pasechnik:2009yc})
\be
\mu_{4}(Q^2)=\mu_{4}(Q^2_0) \, \frac{\tilde{A}^{(1)}_{{\rm MA},d_4,0}(Q^2)}{\tilde{A}^{(1)}_{{\rm MA},d_4,0}(Q^2_0)}\, .
\label{mu4Q2MA} 
\ee

Since we are including very small values of $Q^2$ here, the above representation (\ref{Gpn.OPE}) of the HT contributions is inconvenient.
It is much better to use the so-called ``massive'' representation for the HT part (introduced in Ref. \cite{Teryaev:2013qba}):
\be
\Gamma_1^{p-n}(Q^2)=\left|\frac{g_A}{g_V}\right| \, \frac{1}{6} \, \bigl(1-D_{\rm BS}(Q^2)\bigr) +\frac{\tilde{\mu}_4}{Q^{2}+M^2} \, ,
\label{Gpn.mOPE} 
\ee
where the values of $\tilde{\mu}$ and $M^2$ has been fitted in Refs. \cite{Ayala:2017uzx,Ayala:2018ulm}
in the different analytic QCD models.

The perturbative part has the following form
\be
D_{\rm BS}(Q^2)=\frac{4}{\beta_0} \, a_s\left(1+d_1a_s+d_2a_s^2+d_3a^3_s\right)=
\frac{4}{\beta_0} \, \left(\tilde{a}_{1}+\tilde{d}_1\tilde{a}_2+\tilde{d}_2\tilde{a}_3+
\tilde{d}_3\tilde{a}_4\right),
\label{DBS} 
\ee
where
\be
\tilde{d}_1=d_1,~~\tilde{d}_2=d_2-b_1d_1,~~\tilde{d}_3=d_3-\frac{5}{2}b_1d_2+\bigl(\frac{5}{2}b^2_1-b_2\bigr)\,d_1,
\label{tdi} 
\ee

For $f=3$ case, we have
\be
\tilde{d}_1=1.59,~~\tilde{d}_2=2.51,~~\tilde{d}_3=10.58 \, .
\label{td123} 
\ee

In the MA model, the perturbative part
has the following form
\be
D_{\rm{MA,BS}}(Q^2) =\frac{4}{\beta_0} \, \Bigl(A^{(k)}_{\rm MA,k-1}
+ \sum_{m=2}^{k} \, \tilde{d}_{m-1} \, \tilde{A}^{(k)}_{\rm MA,\nu=m, k-1} \Bigr)\,.
\label{DBS.ma} 
\ee

Moreover, from \cite{Ayala:2018ulm} one can see that in (\ref{Gpn.mOPE})
\be
M^2=0.439,~~\tilde{\mu}_4=-0.082 \, .
\label{M,mu} 
\ee

\subsection{Results}

\begin{figure}[!htb]
\centering
\includegraphics[width=0.58\textwidth]{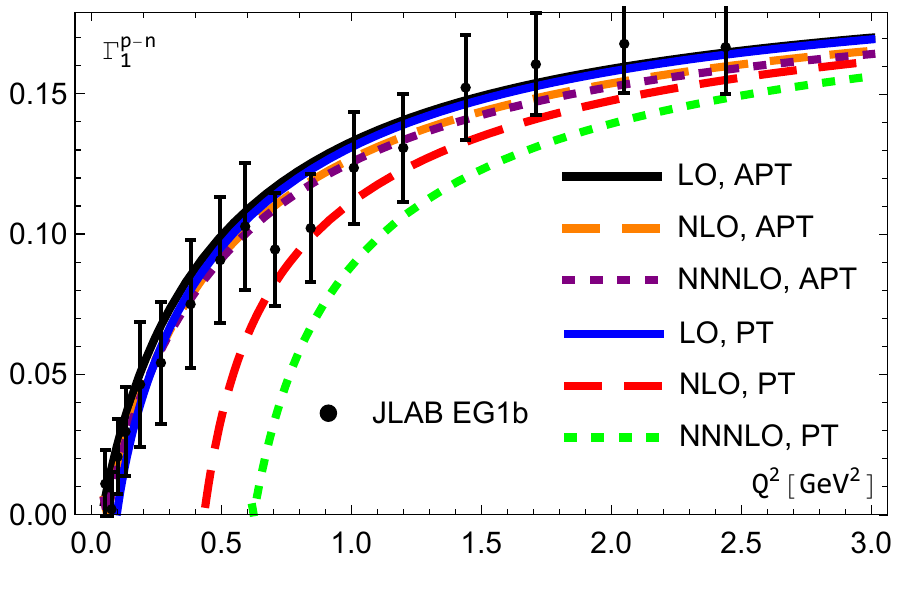}
    \caption{\label{fig:PTHT}
      Results for $\Gamma_1^{p-n}(Q^2)$ in the first, second? and fourth orders of APT and usual PT, obtained with Eqs.
      (\ref{DBS}) and (\ref{DBS.ma}) for the twist-two part. Experimental points are taken from  \cite{Deur}.}
\end{figure}

The results of calculations are shown in Fig. \ref{fig:PTHT}.
Here we use the $Q^2$-independent $M$ and $\tilde{\mu}_4$ values taken from (\ref{M,mu}), as well as the twist-two parts shown in Eqs. (\ref{DBS}) and (\ref{DBS.ma})
for the cases of conventional perturbation theory and APT, respectively.

As can be seen in fig. \ref{fig:PTHT}, the results obtained with conventional coupling constants
are only good at LO and worsen as the order of perturbation theory increases. Good agreement at LO is due to the use of $\Lambda_{\rm LO}$, which is small
(see (\ref{Lambdas})), and therefore the range of $Q^2$ under study is higher than $\lambda^2_{\rm LO}$.
Visually, these results are close to those obtained in Ref. \cite{Khandramai:2011zd}, where the "massive" form (\ref{mu4Q2}) of the twist-four part was also used.
Thus, the use of this "massive" form of the twist-four term (\ref{Gpn.mOPE}) does not improve the results, since at $Q^2 \to \Lambda_i^2$ ordinary coupling constants become
singular and this leads to large and negative results for the contribution (\ref{DBS}) of the twist-two part. As the order of the perturbation theory increases,
ordinary coupling constants become singular for larger and larger values of $Q^2$ (see Fig. \ref{fig:as1352}), so the Björken sum rule tends to negative values as $Q^2$ increases.
Thus, the discrepancy between theory and experiment increases with the order of the perturbation theory.

In the case of MA coupling constants, our results are close to those obtained in Ref. \cite{Ayala:2018ulm}, which is not surprising since we used the parameters
(\ref{M,mu}) taken from \cite{Ayala:2018ulm}.
Moreover, we see that the results based on different orders of perturbation theory are close to each other, in contrast to the case of using ordinary coupling constants.

So, we see that our results for $\Gamma_1^{p-n}(Q^2)$ in the framework of ordinary and MA strong coupling constants are very similar to the results obtained
by \cite{Khandramai:2011zd} and \cite{Ayala:2017uzx,Ayala:2018ulm} respectively.
In the future, we plan to extend our present research to using non-minimal versions of analytic coupling constants (see \cite{CPCCAGC,3dAQCD}) and study the Bjorken sum rule
$\Gamma_1^{p-n}(Q^2)$ within these versions.

\section{Conclusions}

In this paper, we have demonstrated the results obtained in our recent paper \cite{Kotikov:2022sos}. In part, Ref.  \cite{Kotikov:2022sos} contains
$1/L$-expansions of $\nu$-derivatives of the
strong coupling constant $a_s$ expressed as combinations of the $\hat{R}_m$ (\ref{hR_i}) operators applied to the LO coupling constant $a_s^{(1)}$.
Using the same operators to $\nu$-derivatives of LO MA coupling constant $A_{\rm MA}^{(1)}$, four different representations were obtained for $\nu$-derivatives of MA coupling constants,
i.e. $\tilde{A}_{\rm MA,\nu}^{(i)}$, in each $i$-order of perturbation theory: one form contains a combination of polylogarithms; the other form contains expansions
of the generalized Euler $\zeta$-function, and the third one is based on dispersion integrals containing the LO spectral function.
Moreover, in Ref. \cite{Kotikov:2022sos} the fourth representation was also obtained, based on the dispersion integral containing the spectral function of the $i$th order.
All results are presented in \cite{Kotikov:2022sos} up to the 5th order of perturbation theory, where the corresponding coefficients of the QCD $\beta$-function are well known
(see \cite{Baikov:2016tgj,Herzog:2017ohr}). In this paper, we have limited ourselves to only the first three orders in order to exclude the most cumbersome
results obtained for the last two orders of perturbation theory.

High-order corrections are negligible in both asymptotics: $Q^2 \to 0$ and $Q^2 \to \infty$, and are nonzero in a neighborhood of the point $Q^2 =\Lambda^2$.
Thus, they are really only minor corrections to LO MA coupling constant $A_{\rm MA,\nu}^{(1)}(Q^2)$.
This proves the possibility of expansions of high-order coupling constants $A_{\rm MA,\nu}^{(i)}(Q^2)$ in terms of LO-coupling constants $A_{\rm MA,\nu}^{(1)}(Q^2)$, which was done in
Ref. \cite{Bakulev:2010gm}, as well as the possibility of various approximations used in
\cite{Pasechnik:2008th,Pasechnik:2009yc,Khandramai:2011zd,Kotikov:2010bm}.

As can be clearly seen, all the results obtained in Ref. \cite{Kotikov:2022sos} have a compact form and do not contain complex special functions, such as Lambert's $W$-function
\cite{Magradze:1999um}, which appears already in two-loop order as an exact solution of the usual strong coupling constant and which was used to obtain results for MA
coupling constants in \cite{Bakulev:2012sm}.

As an example, following Ref. \cite{Kotikov:2022sos} we considered the Bjorken sum rule and obtained results similar to previous studies in Refs.
\cite{Pasechnik:2009yc,Khandramai:2011zd,Pasechnik:2008th,Ayala:2017uzx,Ayala:2018ulm,Chen:2006tw,Chen:2005tda,Kotikov:2012eq},
because the high order corrections are small in the case of the MA coupling constant.
The results based on the usual perturbation theory do not agree with the experimental data at $Q^2 \leq 1.5$ GeV$^2$ but MA APT leads to good agreement with
the experimental data when we used the "massive" version (\ref{Gpn.mOPE}) for twist-four contributions.

In the future, as was also discussed in \cite{Kotikov:2022sos}, we plan to apply the results odserved here to study the processes of deep inelastic scattering (DIS)
at small $Q^2$ values.
One of the most important applications is the fitting of experimental data for the DIS structure functions $F_2(x,Q^2)$ and $F_3(x,Q^2)$ (see.
\cite{PKK,Shaikhatdenov:2009xd,Kotikov:2015zda,KK2001} and \cite{KKPS1,KPS} respectively). This is one of the main ways to define $\alpha_s(M_Z)$, the strong
coupling constant normalization.
In the studies, we plan to use the $\nu$-derivatives of MA coupling constant $\tilde{A}_{\rm MA,\nu}^{(i)}(Q^2)$ in our approach, and this is indeed possible, because when
fitting we use the Mellin moments of the structure functions (following Ref. \cite{Barker}) and restore the structure functions themselves only
at the very end.
This approach differs from the more popular \cite{NNLOfits} approaches based on the numerical solution of the Dokshitzer-Gribov-Lipatov-Altarelli-Parisi (DGLAP)
\cite{DGLAP} equations.
In the case of using our approach (see \cite{Barker}), the $Q^2$ dependence of the SF moments is known exactly in analytical form (see, for example,
\cite{Buras}): it can be expressed in terms of $\nu$-derivatives $\tilde{A}_{\rm MA,\nu}^{(i)}(Q^2)$, where the corresponding $\nu$-variable becomes
$n$-dependent (here $n$ is the value of the Mellin moment), and the use of $\nu$-derivatives should be crucial.
Above LO, to obtain correct analytic results for the Mellin moments, we plan to use their analytic continuation \cite{KaKo}.

Note that after resumming for large values of the Bjorken variable $x$, the results of the twist-four terms for the structure function $F_2(x,Q^2)$ changed
sign at large $x$ values (see \cite{Kotikov:2022vlo}). Thus, in contrast to the standard analyzes performed in \cite{Kotikov:2010bm}, in the case under consideration,
a part of the twist-four terms should be absorbed into the difference between ordinary and MA coupling constants, similarly, as was done in the
studies \cite{CIKK09,Kotikov:2012sm} carried out for small values of $x$ in the framework of the so-called double asymptotic scaling approach \cite{Q2evo}.

In addition, as the next steps, it is planned to include into consideration the high order terms obtained in the case of more complicates MA coupling constants
(following Refs. \cite{Bakulev:2006ex,Bakulev:2010gm,Ayala:2018ifo,Mikhailov:2021znq}), as well as in case of non-minimal versions of analytic coupling constants
(following links \cite{Cvetic:2006mk,Cvetic:2006gc,Cvetic:2010di,CPCCAGC,3dAQCD}).  For non-minimal versions of analytic coupling constants, only integral representations
(\ref{disptAnuz.ma}) and (\ref{disptAnuz.ma1}) can be used. They, in turn, show the importance of using $\nu$-derived MA coupling constants
(see also the recent paper \cite{Kotikov:2022JETP} and discussions therein). Indeed, in this case it
is necessary to work with the spectral function of the coupling constant $a_s(Q^2)$, and not with the corresponding spectral functions
its $\nu$-powers, the calculation of which requires a very complicated procedure (see \cite{Bakulev:2012sm}).


\section{Acknowledgments}
This work was supported in part by the Foundation for the Advancement
of Theoretical Physics and Mathematics “BASIS”.
 One of us (A.V.K.) thanks the Organizing Committee of the International Workshop on
Elementary Particles and Nuclear Physics (April 24-30, Almaty, Kazakhstan)
for invitation.

\appendix
\def\theequation{A\arabic{equation}}
\setcounter{equation}{0}

\section{Details of evaluation of the fractional derivatives}

Taking the results
(\ref{as}) of the coupling constant $a_s(Q^2)$ we get the following results for the $1/L$-expansion of
its $\nu$-powers:
\be
\left(a^{(1)}_{s,0}(Q^2) \right)^{\nu} = \frac{1}{L^{\nu}_0},~~
\left(a^{(i+1)}_{s,i}(Q^2) \right)^{\nu} = 
\left(a^{(1)}_{s,i}(Q^2) \right)^{\nu} + \sum_{m=2}^i \, \delta^{(m)}_{\nu,i}(Q^2)
\,,~~(i=0,1,2,...)\,,
\label{as_nu}
\ee
where $L_k$ is defined in Eq. (\ref{L}) and
\be \delta^{(2)}_{nu,k}(Q^2) = - \frac{b_1\nu \ln L_k}{L_k^{\nu+1}} ,~~
\delta^{(3)}_{\nu,k}(Q^2) =  \frac{1}{L_k^{\nu+2}} \, \left[b_1^2\left(\frac{\nu+1}{2}\ln^2 L_k-\ln L_k-1\right)+b_2\right]\, ,
\label{ds_nu}
\ee
which is consistent with the expansions made in  
Refs. \cite{BMS1,Bakulev:2006ex}.

The $(\nu-1)$-derivative $\tilde{a}_{\nu}(Q^2)$
is related with the $\nu+l$ $(l=0,1,2,...)$ powers as follows
\be
\tilde{a}_{\nu}(Q^2)=  a_s^{\nu}(Q^2) + k_1(\nu) a_s^{\nu+1}(Q^2) + k_2(\nu) a_s^{\nu+2}(Q^2)
+ O(a_s^{\nu+3}) \, ,
\label{tanu}
\ee
where (see \cite{GCAK})
\be k_1(\nu)=\nu b_1 \, B_1(\nu),~~
k_2(\nu)= \nu(\nu+1) \, \left(b_2 \,  B_2(\nu) + \frac{b_1^2}{2} \, B_{1,1}(\nu) \right),~~
\label{ki}
\ee
with
\be B_1(\nu)=S_1(\nu)-1,~~B_2(\nu)=\frac{\nu-1}{2(\nu+1)},~~B_{1,1}(\nu)= Z_2(\nu+1)-2S_1(\nu)+1
\label{Bi},
\ee
and
\be
Z_2(\nu)=S_1^2(\nu)-S_2(\nu),~~
Z_1(\nu)\equiv S_1(\nu)=\Psi(1+\nu)+\gamma_{\rm E},~~S_2(\nu)=\zeta_2-\Psi'(1+\nu),
\label{si}
\ee
with Euler constant $\gamma_{\rm E}$ and Euler functions $\zeta_2$, $\zeta_3$ and $\zeta_4$.

After some calculations, we get
\be
\tilde{a}^{(1)}_{\nu,0}(Q^2) =\frac{1}{L_0^{\nu}},~~
\tilde{a}^{(i+1)}_{\nu,i}(Q^2) =
\tilde{a}^{(1)}_{\nu,i}(Q^2)+\sum_{m=1}^i C_m^{\nu+m} \,\tilde{\delta}^{(m+1)}_{\nu,i}(Q^2) \, ,
\label{tanu.1b}
\ee
where
\be
\tilde{\delta}^{(m+1)}_{\nu,k}(Q^2) = R_{m,k} \, \frac{1}{L_k^{\nu+m}}
\label{tdelta_mk}
\ee
and
\be R_{1,k}=b_1 \Bigl[\hat{Z}_1(\nu)
  -\ln L_k\Bigr],~~R_{2,k}=b_2 + b_1^2 \Bigl[\ln^2 L_k -2 \hat{Z}_1(\nu+1)\ln L_k + \hat{Z}_2(\nu+1 )\Bigr]\,,
\label{R_i}
\ee
$C_m^{\nu+m}$ is given in Eq. (\ref{tdmp1N}) and 
\be \hat{Z}_1(\nu)=Z_1(\nu)-1,~~\hat{Z}_2(\nu)=Z_2(\nu)-2Z_1(\nu)+1,
\label{Z_123}
\ee
where $Z_i(\nu)$ are defined in Eq. (\ref{si}).

It is convenient to introduce 
the operators $\hat{R}_i$ (\ref{tdmp1N}), which can be obtained as $\hat{R}_i=R_{i,k}\bigl(\ln L_k \to -d/(d\nu)\bigr)$. In this case, we proceed to the results
   (\ref{tdmp1N}) and (\ref{hR_i}) of the main text.

\def\theequation{B\arabic{equation}}
\setcounter{equation}{0}
\section{Alternative
  form for the coupling constants $\tilde{A}^{(i+1)}_{\rm MA,\nu,i}(Q^2)$}

The functions $\zeta(n,-\nu-r -k)$ in (\ref{tRi}) are not so good defined at large $r$ values
and we will replace them the using
the result (\ref{ze_nu.1}) as
\be
\zeta(\nu-r)= -\frac{\Gamma(\nu+r+1)}{\pi(2\pi)^{\nu+r}}\,\tilde{\zeta}(\nu+r+1),~~\tilde{\zeta}(\nu+r+1)=\sin\left[\frac{\pi}{2}(\nu+r)\right] \, \zeta(\nu+r+1) \, .
  \label{ze_nu.1N}
\ee

After some calculations we have
\be
\tilde{\delta}^{(m+1)}_{{\rm MA},\nu,k}(Q^2)=
\frac{1}{\Gamma(\nu+m)} \, \sum_{r=0}^{\infty} \frac{\Gamma(\nu+r+m)}{\pi(2\pi)^{\nu+r+m-1}} \, Q_m(\nu+r+m) 
\, \frac{(-L_k)^r}{r!}\,,
\label{tdAmanNew}
\ee
where
\bea
&&Q_1(\nu+r+1)=b_1\Bigl[\tilde{Z}_1(\nu+r)\tilde{\zeta}(\nu+r+1)+\tilde{\zeta}_1(\nu+r+1)\Bigr], \nonumber \\
&&Q_2(\nu+r+2)=b_2\tilde{\zeta}(\nu+r+2) + b_1^2\Bigl[\tilde{\zeta}_2(\nu+r+2) + 2\tilde{Z}_1(\nu+r+1)\tilde{\zeta}_1(\nu+r+2) \nonumber \\
  &&\hspace{0.5cm} + \tilde{Z}_1(\nu+r+1)  \tilde{\zeta}(\nu+r+2)
  \Bigr],
\label{Qi}
\eea
with (see also (\ref{si}) and  (\ref{Z_123}))
\bea
&&
\overline{Z}_2(\nu)=\overline{S}_1^2(\nu)-S_2(\nu),~~
\overline{Z}_1(\nu)\equiv \overline{S}_1(\nu)=\Psi(1+\nu)+\gamma_{\rm E} -\ln(2\pi),
\nonumber \\&&
S_2(\nu)=\zeta_2-\Psi'(1+\nu),
\label{osi}
\eea
and
\be \tilde{Z}_1(\nu)=\overline{Z}_1(\nu)-1,~~\tilde{Z}_2(\nu)=\overline{Z}_2(\nu)-2\overline{Z}_1(\nu)+1,~~
\, .
\label{oZ_123}
\ee

Moreover we use here
\be
\tilde{\zeta}_k(\nu)=  \frac{d^k}{(d\nu)^k} \, \tilde{\zeta}(\nu) \, .
   \label{tzetaknu}
\ee

Using the definition of $\tilde{\zeta}(\nu)$ in (\ref{ze_nu.1N}), we have
\bea
&&\tilde{\zeta}_1(\nu+r+1)=\sin\left[\frac{\pi}{2}(\nu+r)\right] \, \zeta_1(\nu+r+1) + \frac{\pi}{2} \,
\cos\left[\frac{\pi}{2}(\nu+r)\right] \, \zeta(\nu+r+1) \, ,\nonumber \\
&&\tilde{\zeta}_2(\nu+r+1)=\sin\left[\frac{\pi}{2}(\nu+r)\right] \, \left(\zeta_2(\nu+r+1) - \frac{\pi^2}{4}\,\zeta(\nu+r+1)\right)
\nonumber \\ && \hspace{1cm}
+ \frac{\pi}{2} \,
\cos\left[\frac{\pi}{2}(\nu+r)\right] \, \zeta_1(\nu+r+1) \, ,
\label{tzeta1234}
\eea
where $\zeta_k(\nu)$ are given in Eq. (\ref{zetaknu}) of the main text.

So, we can rewrite
the results (\ref{tdAmanNew}) with
\bea
&&Q_{m}(\nu+r+m)=\sin\left[\frac{\pi}{2}(\nu+r+m-1)\right] Q_{ma}(\nu+r+m) \nonumber\\
&&+ \frac{\pi}{2} \,
\cos\left[\frac{\pi}{2}(\nu+r+m-1)\right] Q_{mb}(\nu+r+m)\, ,
\label{Qmab}
\eea
where
\bea
&&Q_{1a}(\nu+r+1)=b_1\Bigl[\tilde{Z}_1(\nu+r)\zeta(\nu+r+1)+\zeta_1(1,\nu+r+1)\Bigr],~~
\nonumber \\ &&Q_{1b}(\nu+r+1)=b_1\zeta(\nu+r+1), \nonumber \\
&&Q_{2a}(\nu+r+2)=b_2\zeta(\nu+r+2) + b_1^2\Bigl[\zeta_2(\nu+r+2) + 2\tilde{Z}_1(\nu+r+1)\zeta_1(\nu+r+2) \nonumber \\
  &&\hspace{0.5cm} + \left(\tilde{Z}_1(\nu+r+1) - \frac{\pi^2}{4}\right)
  \zeta(\nu+r+2)
  \Bigr], \nonumber \\
&&Q_{2b}(\nu+r+2)=2b_1^2\Bigl[\tilde{Z}_1(\nu+r+1)\zeta(\nu+r+2)+\zeta_1(\nu+r+2)\Bigr]\,.
\label{Qiab}
\eea

\end{document}